\documentclass[prc,aps,showpacs,twocolumn]{revtex4}

\usepackage{graphicx}

\begin{document}

\title{Prediction and formation mechanism of triaxial superdeformed 
       nuclei for $A\sim 80$}
\author{Caiwan Shen$^{1,2}$, Y. S. Chen$^{2,3}$, E. G. Zhao$^{3}$}
\affiliation{
$^{1}$INFN-LNS, Via S. Sofia 44, I-95123 Catania, Italy \\
$^{2}$China Institute of Atomic Energy, P.O.Box 275(18), Beijing, 102413, China \\
$^{3}$Institute of Theoretical Physics, Academia Sinica, 
      P.O.Box 2735, Beijing, 100080, China }

\begin{abstract}
The three dimensional Total Routhian Surface (TRS) calculations
are carried out for 64 nuclei between $70\leq A\leq 90$ to find
triaxial superdeformed nuclei. Total of 12 nuclei were predicted
to have triaxial superdeformation in which the neutron rotational
energy plays a key role and the neutron shell energy plays
additional role in the formation of triaxial superdeformed nuclei.
\end{abstract}

\maketitle

\section{Introduction}

Usually, the shape of a deformed nucleus is supposed as a
ellipsoid with small hexadecapole deformation. In nilsson model,
the frequency of harmonic oscillator is described as
\begin{equation}
\omega _{k}=\omega _{0}[1-\frac{2}{3}\varepsilon _{2}\cos (\gamma
+k\frac{ 2\pi }{3})],\rm{ \ \ \ \ \ \ \ \ }k=1,2,3,  \label{eq1}
\end{equation}
where $\varepsilon _{2}$ is the quadrupole deformation parameter
and $\gamma $ is the triaxial deformation parameter. Let the half
axis of $x$, $y$, $z$ are $a$, $b$, $c$, respectively, then from
$a\omega _{1}=b\omega _{2}=c\omega _{3}$, we get the relations
between $a$, $b$, $c$ and $ \varepsilon _{2}$, $\gamma $:
\begin{eqnarray}
\varepsilon _{2}
&=&\frac{\sqrt{9(bc+ac-2ab)^{2}+27(bc-ac)^{2}}}{2(ab+ac+bc)}
,  \label{eq2} \\
\gamma &=&\arctan \left( \frac{\sqrt{3}(b-a)}{a+b-2ab/c}\right) ,
\label{eq3}
\end{eqnarray}
and
\begin{eqnarray}
a &=&Rf[\varepsilon _{2}\cos \gamma +\sqrt{3}\varepsilon _{2}\sin \gamma
+3]^{-1},  \nonumber \\
b &=&Rf[\varepsilon _{2}\cos \gamma -\sqrt{3}\varepsilon _{2}\sin \gamma
+3]^{-1},  \label{eq4} \\
c &=&Rf[3-2\varepsilon _{2}\cos \gamma ]^{-1},  \nonumber
\end{eqnarray}
here $f=[-2\varepsilon _{2}^{3}\cos (3\gamma )-9\varepsilon
_{2}^{2}+27]^{1/3}$ and $R$ is the radius of a sphere whose volume
is equal to the ellipsoid volume. For example, when
$a:b:c=1:1:2$(axial ellipsoid), the $\varepsilon _{2}$ and $\gamma
$ will take the volume of $0.6$ and $ 0^{\circ }$, respectively.
Such nuclei, i.e., $c:a\sim 2:1$ and $a\sim b$, is called axial
superdeformed nuclei. While $a\neq b$, the triaxial deformation
$\gamma $ will not be zero. For instance, if $a:b:c=3:4:6$, then
$(\varepsilon _{2},\gamma )$ will be $(0.577,30^{\circ })$. Such
nuclei which has large $\gamma $ deformation and large quadrupole
deformation is called triaxial superdeformed (TSD) nuclei. From
Eq.(\ref{eq3}) we can see that $\gamma $ is sensitive to the
difference of $a$ and $b$.

Until now, many data for axial superdeformed nuclei are
accumulated but the data for triaxial superdeformed nuclei are
quite few. Only five nuclei, $ ^{163}$Lu, $^{165}$Lu, $^{167}$Lu,
$^{171}$Ta, $^{86}$Zr[1-5] were identified experimently to have
triaxial superdeformation. the triaxial behavior of $ ^{163}$Lu
has been further confirmed experimentally by the discovery of the
wobbling mode \cite{od01}. However, the triaxiality of most these
TSD nuclei can only be explained by the theoretical calculations.
Among these five discovered TSD nuclei,  four are located in
$A\sim 160$ region and only one is in the $A\sim 80$ region. The
prediction of triaxial superdeformed nuclei in $A\sim 160$ region
has been done[6]. From that work several other nuclei besides the
four discovered, were predicted to have triaxial superdeformation
for proton configuration $ [660]1/2$. It is also pointed out that
the shapes may co-existences for other curtain configurations and
a nucleus may have triaxial superdeformation in different q.p.
configurations.

The TSD nuclei in \ $A\sim 80$, $^{86}$Zr, was discovered in 1998.
It does not seems like that TSD $^{86}$Zr is a accident appeared
in the $A\sim 80$ region. Other TSD nuclei must exist in this
region. In this paper, we attempt to give the prediction of the
TSD nuclei near $A\sim 80$ by TRS (Total Routhian Surface)
calculations and their formation mechanism . Section 2 includes
brief description of the three-dimensional TRS theory which is
used to determine the nuclear deformation; Section 3 includes the
prediction of TSD nuclei in $A\sim 80$ region; Section 4 includes
the discussion of the formation mechanism of the TSD nuclei;
Section 5 gives a summary.

\section{A brief description of the TRS model}

The hamiltonian of quasi-particles moving in a quadrupole deformed potential
rotating around the $x$-axis with a frequency $\omega $ may be written as
\begin{equation}
H^{\omega }=H_{{\rm s.p.}}(\varepsilon _{2},\varepsilon _{4},\gamma
)-\lambda N+\Delta (P+P^{+})-\omega J_{x},  \label{eq5}
\end{equation}
where $H_{{\rm s.p.}}$ denotes the deformed hamiltonian of single
particle motion, the second term on the right hand side is the
chemical potential, the third term is the pairing interaction and
the last term stands for the Coriolis forces. The
modified-harmonic-oscillator (MHO) potential with the parameters
$\kappa $ and $\mu $ for the mass region taken from Ref.[7] is
employed in the present calculation. The pairing-gap parameter is
determined empirically by $\Delta =0.9\Delta _{\rm{{\rm o.e}.}}$,
and $\Delta _{{\rm o.e.}}$ is taken from experimental odd-even
mass difference [8]. As an approximation, we did not take the
deformation and rotation dependence of pairing into account.

The total routhian surface, namely the total energy in the rotating frame as
a function of the $\varepsilon _{2}$, $\gamma $, and $\varepsilon _{4}$, of
a $(Z,N)$ nucleus for a fixed quasi-particle configuration c.f. may be
calculated by
\begin{eqnarray}
E^{{\rm c.f.}}(\varepsilon _{2},\varepsilon _{4},\gamma ;\omega )
&=&E_{{\rm ld}}(\varepsilon _{2},\varepsilon _{4},\gamma )+E_{{\rm
corr}}(\varepsilon
_{2},\varepsilon _{4},\gamma ;\omega =0)  \nonumber \\
&&+E_{{\rm rot}}(\varepsilon _{2},\varepsilon _{4},\gamma ;\omega
)+\sum\limits_{i\in {\rm c.f.}}e_{i}^{\omega }(\varepsilon _{2},\varepsilon
_{4},\gamma ),  \label{eq6}
\end{eqnarray}
Where $E_{{\rm ld}}$ is the liquid-drop model energy[9], $E_{{\rm
corr}}$ is the quantal-effect correction to the energy, which
includes both the shell[10] and pairing corrections[11]. The
collective rotational energy $E_{ {\rm rot}}$ may be calculated
microscopically as the energy difference between the expectation
values of $H^{\omega }$ with and without rotation, by using the
wave function for the quasi-particle vacuum configuration[12]. The
last term of Eq.(2) is the sum of quasi-particle energies
belonging to the configuration c.f., which generates the
deformation drive. All of the term in Eq.(2) depend on the $(Z,N)$
numbers which are not written explicitly. The equilibrium
deformations of nucleus may be calculated by minimizing the total
routhian energy of Eq.(2) with respect to $\varepsilon _{2}$,
$\varepsilon _{4}$, and $\gamma $. Here we put the hexadicapole
deformation $\varepsilon _{4}$ as a free parameter in order to get
better results.

In the real process of minimizing the total routhian, we minimize
the total routhian as a function of $\varepsilon _{4}$ for each
point $(\varepsilon _{2},\gamma )$ and get two surface $E^{{\rm
c.f.}}(\varepsilon _{2},\gamma )$ and $\varepsilon
_{4}(\varepsilon _{2},\gamma )$ first. Then from the surface of
$E^{{\rm c.f.}}(\varepsilon _{2},\gamma )$ we can find the minimum
$E_{\min }^{{\rm c.f.}}$ and the corresponding $\varepsilon
_{2{\rm { min}}}$ and $\gamma _{\min }$. From the $\varepsilon
_{4}(\varepsilon _{2},\gamma )$, $\varepsilon _{2{\rm {min}}}$ and
$\gamma _{\min }$, we may determine the value of $\varepsilon
_{4\min }$. From these steps above, we may find the equilibrium
deformation $\varepsilon _{2\min }$, $\varepsilon _{4\min }$ and
$\gamma _{\min }$ which possibly exists in the nucleus.

\section{The prediction of the triaxial superdeformed nuclei}

Before the calculation of $A\sim 80$ nuclei, the TRS method was
checked and compared with the discovered TSD nuclei, $^{86}$Zr.
Our calculated result, $ (\varepsilon _{2},\gamma )=(0.455,16.8)$,
is very coincident with the result in Ref.[5] , indicating the
reliability of the method for the $A\sim 80$ mass region.

\begin{figure}[hbt]
\includegraphics[width=8cm]{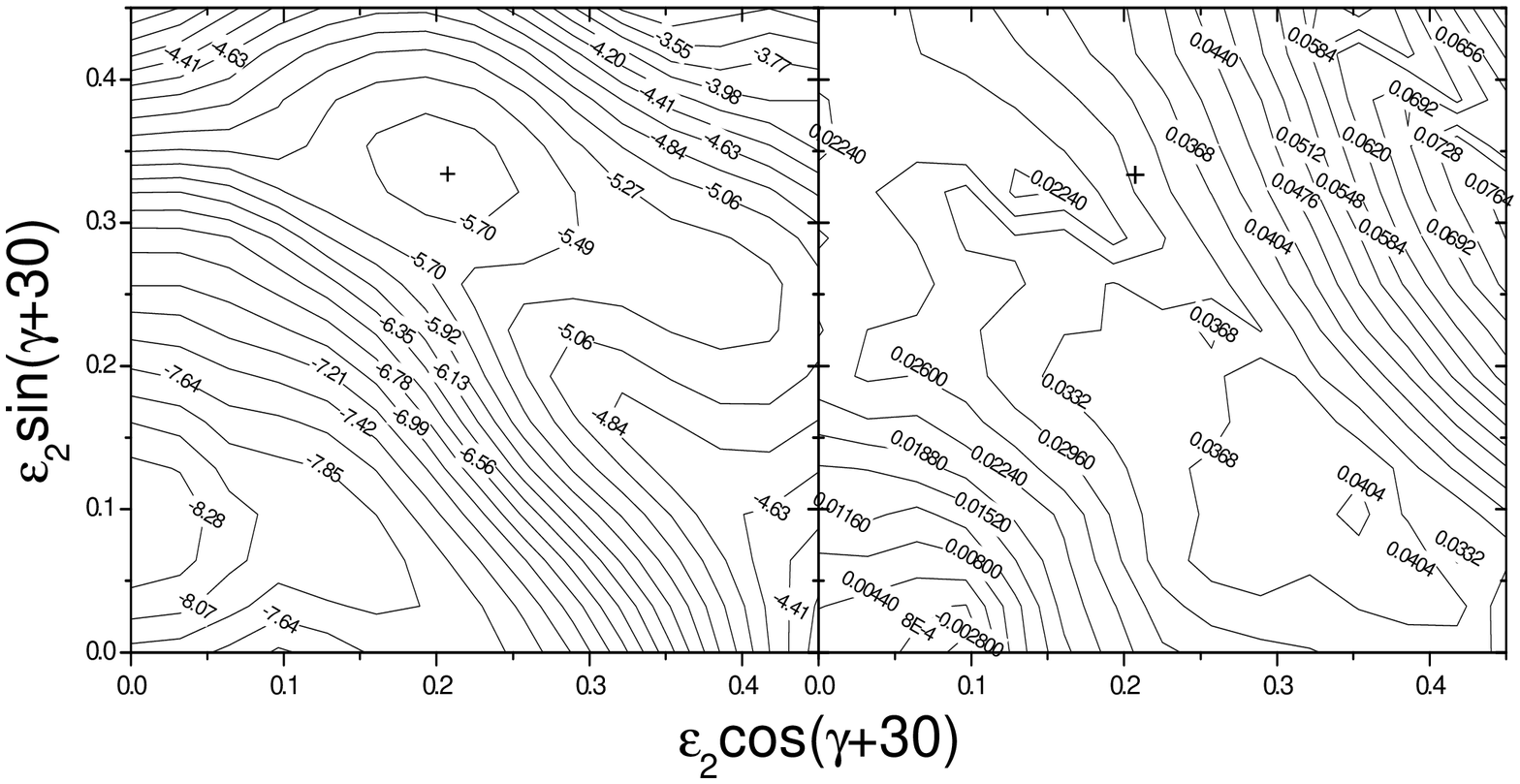}
\label{fig1}
\caption{ The shape determination of $^{80}$Kr. (a) the counter plot of total
routhian. The unit of the number in the surface is MeV. ``+'' indicates a local
minimum which deformation is $(\varepsilon _{2},\gamma )=(0.393,28.8^{\circ
})$. (b) the counter plot of $\varepsilon _{4}$. The hexadecapole
deformation at symbol ``+'', which has the same position as ``+'' in (a), is
0.030.}
\end{figure}

In the following, the progress that used to determine the
deformation of a nuclei will be described detailly with the
example of $^{80}$Kr. From the fact that the $\gamma $-ray energy
within a superdeformed bands in $A\sim 80$ is much higher than
that in $A\sim 160$, we get that the superdeformed nuclei in
$A\sim 80$ rotates much faster than that in $A\sim 160$ because
the rotational frequency, $\omega $, is approximately the half of
the $\gamma $ -ray energy. Thus, when we predict the shape of a
nucleus in $A\sim 80$, the $\omega $ must be larger. In this
paper, we fixed the $\omega $ as $ 0.1\omega _{0}$, where the
$\omega _{0}=41/\sqrt[3]{A}$ MeV. In the three-dimensional
calculation, the $\varepsilon _{4}$ from $-0.04$ to $0.10$ is
divided to 11 points. The total routhian energy in each
$(\varepsilon _{2}\cos (\gamma +30^{\circ }),\varepsilon _{2}\sin
(\gamma +30^{\circ }))$ point will be minimized with respect to
the corresponding 11 points. Fig.1(a) shows a contour plot of the
total routhian surface in which each point corresponds to the same
$\omega $ but different $\varepsilon _{4}$. In the Fig.1(a), there
is a local minimum marked by ``+'' which has the deformation
$(\varepsilon _{2},\gamma )=(0.393,28.8^{\circ })$. The
hexadecapole parameter $ \varepsilon _{4}$, corresponding to the
local minimum in Fig.1(a), is determined by Fig.1(b) which is the
counter plot of $\varepsilon _{4}$. The $ \varepsilon _{4}$ in
each grid point in Fig.1(b)\ is got from the minimization of the
total routhian against $\varepsilon _{4}$. The symbol + in
Fig.1(b) which corresponds to the minimum in Fig.1(a) has the $
\varepsilon _{4}$ as 0.030. Thus, the deformation of $^{80}$Kr at
$\omega =0.1\omega _{0}$ has been determined as $(\varepsilon
_{2},\gamma ,\varepsilon _{4})=(0.393,28.8^{\circ },0.030)$.
During the calculation, we do not add the quasi-particle energy
(the last item in Eq.(\ref{eq6})) to the total routhian energy
because for so high rotational frequency, one or two pairs of
particle has been broken and it has been automatically included in
the rotational energy part (see Sec.IV for details).

\begin{table}
\caption {\label{table1}
Predicted TSD nuclei between $70\leq A\leq 90$ for rotational
frequency of $0.1\hbar\omega _{0}$ under the selection of
$\varepsilon _{2}>0.35$ and $10^{\circ}\leq \gamma\leq 50^{\circ}$ }
\begin{tabular}{cccc|cccc}
\hline\hline Nuclei & $\varepsilon _{2}$ & $\gamma $ &
$\varepsilon _{4}$ & Nuclei & $ \varepsilon _{2}$ & $\gamma $ &
$\varepsilon _{4}$ \\  \hline $^{72}$Ni & 0.448 & 12.6$^{\circ }$
& 0.048 & $^{80}$Kr & 0.394
          & 28.6$^{\circ }$ & 0.030 \\
$^{74}$Ni & 0.393 & 21.3$^{\circ }$ & 0.023 & $^{86}$Zr & 0.471
          & 19.0$^{\circ }$ & 0.043 \\
$^{76}$Zn & 0.404 & 21.3$^{\circ }$ & 0.029 & $^{88}$Mo & 0.500
          & 14.7$^{\circ }$ & 0.055 \\
$^{76}$Ge & 0.403 & 30.7$^{\circ }$ & 0.027 & $^{90}$Mo & 0.375
          & 40.0$^{\circ }$ & 0.037 \\
$^{78}$Se & 0.396 & 32.7$^{\circ }$ & 0.027 & $^{90}$Ru & 0.483
          & 23.1$^{\circ }$ & 0.045 \\
$^{80}$Se & 0.351 & 36.1$^{\circ }$ & 0.022 &  &  &  & \\
\hline\hline
\end{tabular}
\end{table}

Following the step described above, we analyzed all of the $\beta
$ stable even-even nuclei between $70\leq A\leq 90$, totally 64
nuclei. The predicted TSD nuclei are listed on Table I. In this
paper, we call the deformation with $\varepsilon _{2}>0.35$ and
$10^{\circ }\leq \gamma \leq 50^{\circ }$ as triaxial
superdeformation. So, in Table 1, only the TSD nuclei under the
deformation condition, $\varepsilon _{2}>0.35$ and $10^{\circ
}\leq \gamma \leq 50^{\circ }$, are listed.

\begin{figure}
\includegraphics[width=8cm]{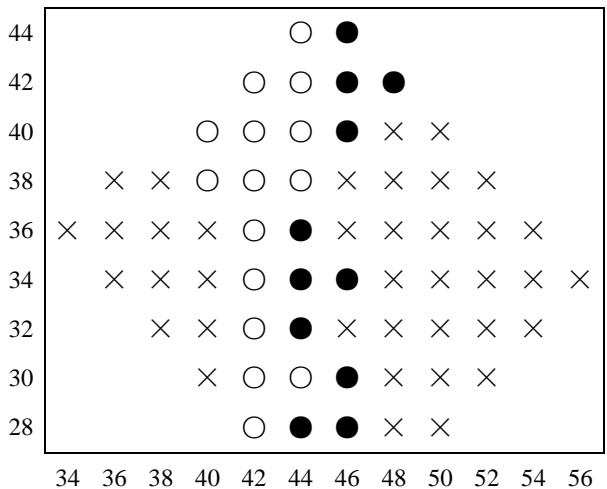}
\label{fig2}
\caption{ The prediction of triaxial superdeformed nuclei. The solid circle,
open circle and cross represents the predicted triaxial, axial,
non-superdeformed nuclei, respectively. It is shown that most of predicted
TSD nuclei are located in $N=44,46$. }
\end{figure}

The location of the predicted nuclei among the nuclei between $70\leq A\leq
90$ is shown in \ Fig.2. In this figure, solid circles represent the
predicted TSD nuclei, open circles axial SD nuclei, cross symbols the nuclei
in which we did not find superdeformation. \ A obvious regular in this
figure is that when $N=42,44,46$, most of the nuclei have superdeformation.
Especially, for nuclei of $N=44,46$, most of them have triaxial
superdeformation. Apparently, the neutron properties control the formation
of axial superdeformation and triaxial superdeformation. How and why do the
neutron properties control the formation mechanism of TSD nuclei?

\section{Formation mechanism of TSD nuclei in $A\sim 80$}

In Fig.2, it is obvious that the neutron numbers of the most
predicted TSD nuclei are 44 and 46. Only $^{90}$Mo is an
exception. This phenomenon indicates that the neutron property
governs the formation of TSD nuclei.

\begin{figure}
\includegraphics[width=8cm]{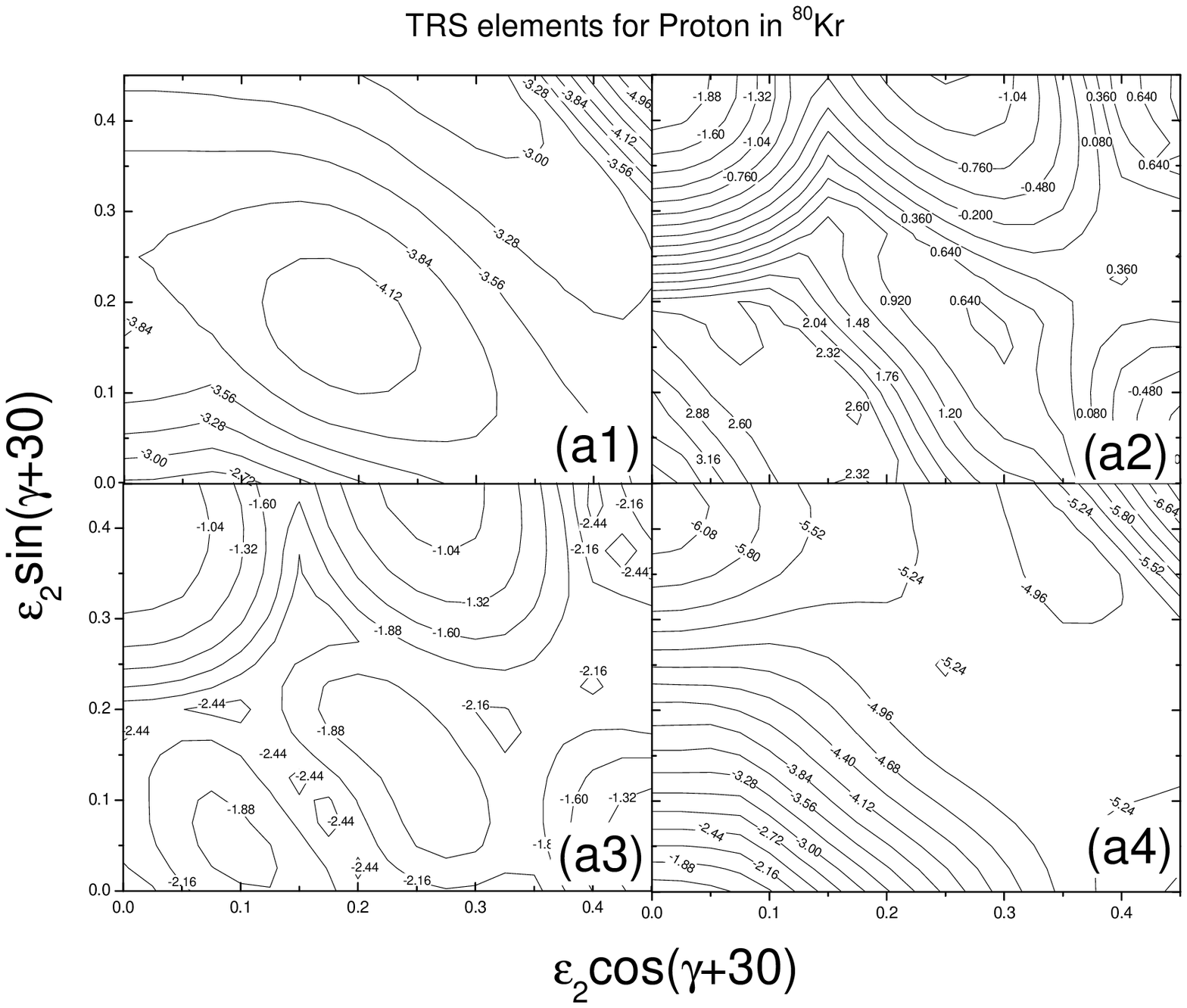} \\
\includegraphics[width=8cm]{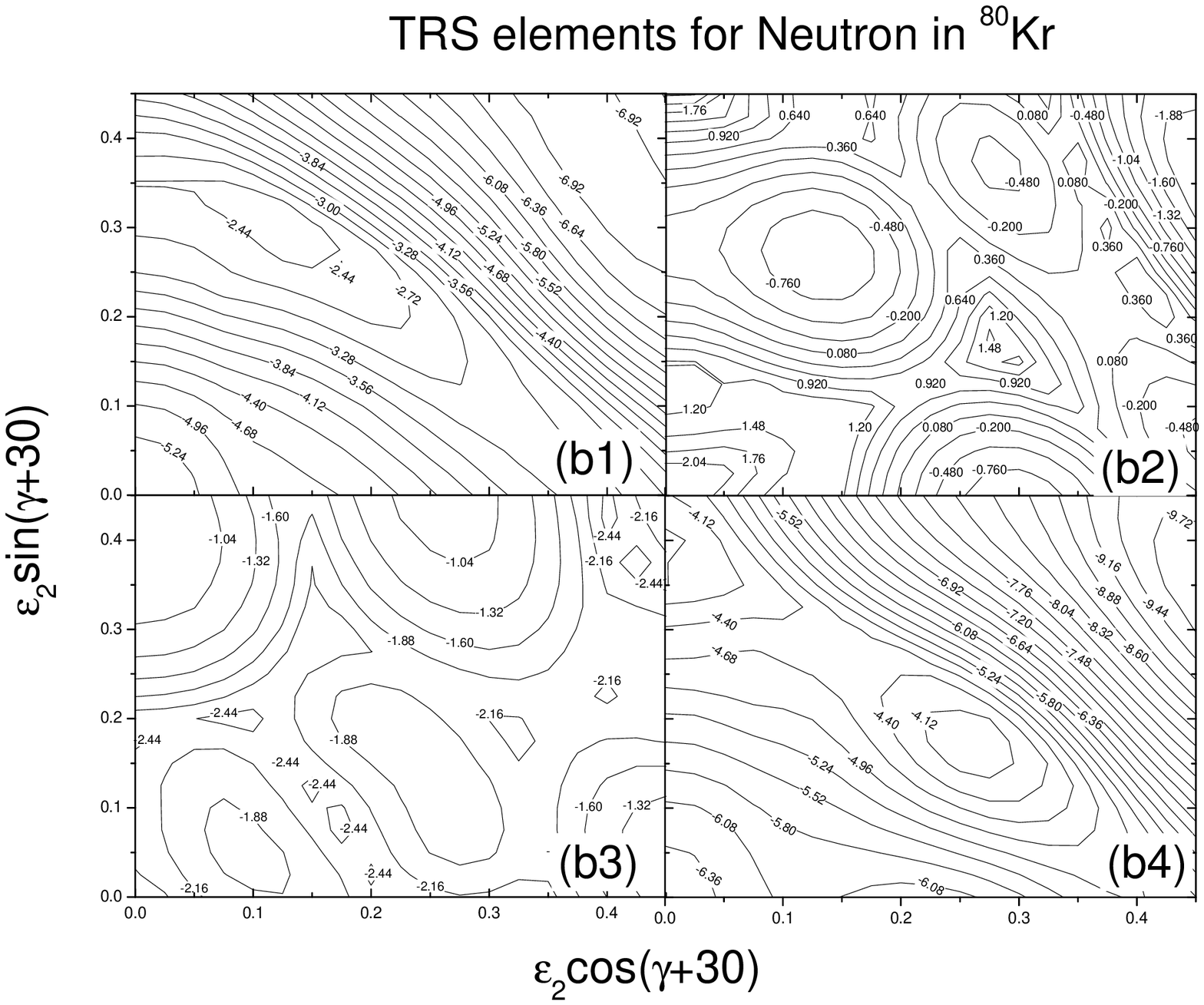}
\label{fig3}
\caption{ The formation mechanism of $^{80}$Kr. (a) and (b) are for protons and
neutrons, respectively. (a1), (b1) is for rotating energy, (a2), (b2) for
shell correction energy, (a3), (b3) for pairing energy, (a4), (b4) is the
sum of the previous three item. It is shown that neutron rotating energy
plays a key role in the formation of predicted TSD $^{80}$Kr. See text for
details.}
\end{figure}

\begin{figure}
\includegraphics[width=8cm]{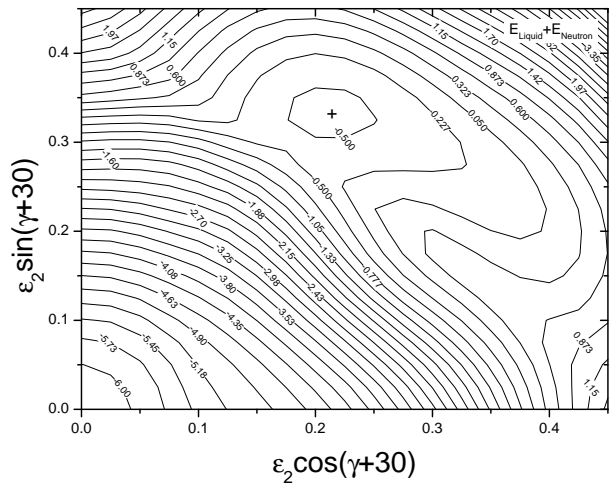}
\label{fig4}
\caption{ The counter plot of the sum of liquid drop energy, neutron rotating
energy, shell correction energy and pair correction energy. This plot shows
the local minimum marked by ``+'' which is close to the local minimum in
Fig.1(a).}
\end{figure}

In order to discuss the mechanism detailly, the $^{80}$Kr, predicted to have
triaxial superdeformation, is selected. According to the Eq.(\ref{eq6}),
most part of the total routhian energy, $E_{\rm{rot}}$, $E_{\rm{shell}}$%
, $E_{\rm{pair}}$ and $E_{\rm{sum}}$($=$ $E_{\rm{rot}}+E_{\rm{shell}%
}+E_{\rm{pair}}$) are plotted in Fig.3. Fig.3(a) and Fig.3(b) show
the TRS elements of proton and neutron in $^{80}$Kr, respectively.
The energy scale of the contour lines is 0.28MeV.

In Fig.3(a1), the surface of proton rotational energy is flat and
the deformation of the local minimum is small. Therefore, the
proton rotating energy cannot affect the formation of TSD shape.
In Fig.3(a2), although the proton shell correction energy has two
deep minimums, the proton pairing correction energy, shown in
Fig.3(a3), has two high peeks near the minimum in Fig.3(a) and
canceled the minimum of shell correction energy. The sum of the
three type of energy, $E_{\rm{sum(p)}}$, shown in Fig.3(a4), is
flat in the center part of the surface. This means that the proton
properties in $ ^{80}$Kr is helpless to form TSD nuclei.

However, the neutron properties, shown in Fig.3(b), is different
from Fig.3(a). The neutron rotational energy, shown in Fig.3(b1)
decreases sharply with increasing large $\varepsilon_2$
deformation and therefore has a strong driving effect towards
large elongation deformation. In Fig.3(b2), the neutron shell
correction has two minimums but were canceled by the pairing
energy shown in Fig.3(b3). Thus, the driving force in large
quadrupole deformation is remained. Summing the three neutron
parts of total routhian energy, $E_{\rm{sum(n)}}$, we obtain
Fig.3(b4). This figure is most similar to Fig.3(b1), having small
driving effect to spherical deformation and large driving effect
to large quadrupole deformation. This is very important for
$^{80}$Kr to form TSD shape. Fig.4 shows the sum of liquid drop
energy and $E_{\rm{sum(n)}}$. A large quadrupole and triaxial
minimum appeared on this surface. Since the $E_{\rm{sum(p)}}$ is
flat in this region, the minimum shown in Fig.4  exists also in
the total routhian surface, see Fig.1. In the formation of TSD
shape, the rotational energy plays a crucial role. Because the
neutron shell correction energy also decreases sharply in large
deformation, it also has the additional role to form TSD shape.

To confirm that the reason is also effective for other nuclei in
$A\sim 80$ region, we also analyzed $^{78}$Se which is predicted
to have TSD shape and $ ^{86}$Kr which is predicted to have no TSD
shape. The results supported our analysis of the formation
mechanism of TSD nuclei that the rotational energy plays a key
role and neutron shell energy plays an additional role in the
formation of TSD nuclei.

\begin{figure}[hbt]
\includegraphics[width=85mm]{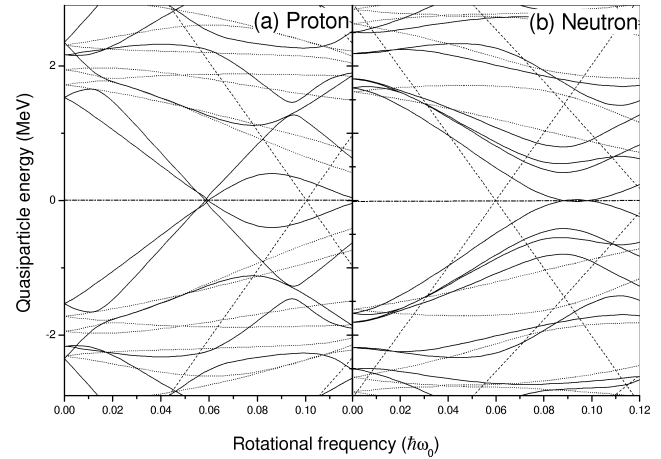}
\label{fig5}
\caption{ The quasipartical energy for protons (a) and
neutrons (b) in $^{80}$ Kr. The following convection is used:
solid lines: ($\pi =+,\alpha =+\frac{1 }{2}$), dotted lines: ($\pi
=+,\alpha =-\frac{1}{2}$), dash-dotted lines: ($ \pi =-,\alpha
=+\frac{1}{2}$) and dashed lines: ($\pi =-,\alpha =-\frac{1}{2}$).}
\end{figure}

It has been pointed out that the rotational energy is the
difference between the expectation value of $H^{\omega }$
(Eq.(\ref{eq5})) with and without rotation. When the rotational
frequency is high, the pair of protons and/or neutrons will be
broken and their angular momentum alignment will affect the
rotational energy. In order to see the  broken pair of protons and
neutrons, the calculated q.p. routhians are presented in Fig.5(a)
for protons and Fig.5(b) for neutrons. In Fig.5(a), [431]3/2 orbit
crosses with the [440]1/2 orbit at $ \omega =0.082\hbar \omega
_{0}$ for protons, while\ in Fig.5(b), \ [420]1/2
orbit crosses with the [413]7/2 orbit for both $\alpha =\pm \frac{1}{2}$ in $%
\omega =0.09\hbar \omega _{0}$ for neutrons. Therefore, When
$^{80}$Kr rotates at $\omega =0.10\hbar \omega _{0}$, there is one
proton pair and two neutron pairs are broken and this will take
effect on the rotational energy.

Compared with the analysis of TSD nuclei in $A\sim 160$ region,
the quadrupole deformation parameters of predicted TSD nuclei in
$A\sim 80$ region are larger than those in $A\sim 160$ region. And
also, the formation mechanism is different between the two
regions. In $A\sim 160$ region, the neutron shell correction
energy controls the formation of TSD nuclei, while in $A\sim 80$
region, rotating energy, which include the effect of q.p. angular
momentum alignments, controls the formation of TSD nuclei.

\section{Summary}

In summary, \ fixing the rotational frequency $\omega $ as
$0.1\hbar \omega _{0}$,11 nuclei are predicted to have triaxial
superdeformation under the condition of $\varepsilon _{2}\geq
0.35$ and $10^{\circ }\leq \gamma \leq 50^{\circ }$ by the three
dimensional TRS calculation in $A\sim 80$ region. Most of these
TSD nuclei are located in the $N=44,46$ region, only $^{90}$Mo is
a exceptional. By analyzing the formation mechanism of TSD nuclei
in the $A\sim 80$ region, we find that the neutron rotational
energy which includes the contribution of quasiparticle angular
momentum alignment plays a key role to form TSD nuclei and the
neutron shell energy plays additional role.

{\bf Acknowledgements}

This work is supported by the NNSF of China under Grant No.s
10075078, 19935030 and 10004701, and the MSBRDP of China under
Grant No. G20000774.

\end{document}